\documentclass[10pt,journal,compsoc]{IEEEtran}

\usepackage{cite}

\usepackage[pdftex]{graphicx}
   \DeclareGraphicsExtensions{.pdf,.png}

\usepackage{algorithmicx}
\usepackage[Algorithm,ruled]{algorithm}
\usepackage{algpseudocode}

\usepackage[table,xcdraw]{xcolor}

\begin{document}
\title{High-Level Programming Abstractions for Distributed Graph Processing}

\author{Vasiliki~Kalavri, Vladimir~Vlassov, and~Seif~Haridi
\IEEEcompsocitemizethanks{\IEEEcompsocthanksitem V. Kalavri, V. Vlassov, and Seif Haridi are with KTH Royal Institute of Technology, Stockholm, Sweden\protect\\
E-mail: kalavri@kth.se, vladv@kth.se, haridi@kth.se}
}

\IEEEtitleabstractindextext{
\begin{abstract}
Efficient processing of large-scale graphs in distributed environments has been an increasingly popular topic of research in recent years. Inter-connected data that can be modeled as graphs arise in application domains such as machine learning, recommendation, web search, and social network analysis. Writing distributed graph applications is inherently hard and requires programming models that can cover a diverse set of problem domains, including iterative refinement algorithms, graph transformations, graph aggregations, pattern matching, ego-network analysis, and graph traversals. Several high-level programming abstractions have been proposed and adopted by distributed graph processing systems and big data platforms. 
Even though significant work has been done to experimentally compare distributed graph processing frameworks, no qualitative study and comparison of graph programming abstractions has been conducted yet. In this survey, we review and analyze the most prevalent high-level programming models for distributed graph processing, in terms of their semantics and applicability. We identify the classes of graph applications that can be naturally expressed by each abstraction and we also give examples of applications that are hard or impossible to express. We review 34 distributed graph processing systems with respect to their programming abstractions, execution models, and communication mechanisms. Finally, we discuss trends and open research questions in the area of distributed graph processing.
\end{abstract}}

\maketitle

\IEEEraisesectionheading{\section{Introduction}\label{sec:introduction}}
\IEEEPARstart{G}{raphs}  are immensely useful data representations, vital to diverse data mining applications. Graphs capture relationships between data items, like interactions or dependencies, and their analysis can reveal valuable insights for machine learning tasks, anomaly detection, clustering, recommendations, social influence analysis, bioinformatics, and other application domains.

The rapid growth of available datasets has established distributed shared-nothing cluster architectures as one of the most common solutions for data processing. Vast social networks with millions of users and billions of user-interactions, web access history, product ratings, and networks of online game activity are a few examples of graph datasets that might not fit in the memory of a single machine. Such massive graphs are usually partitioned over several machines and processed in a distributed fashion. However, coping with huge data sizes is not the sole motivation for distributed graph processing. Graphs rarely appear as \emph{raw} data; they are most often derived by transforming other datasets into graphs. Data entities of interest are extracted and modeled as graph nodes and their relationships are modeled as edges. Thus, graph representations frequently appear in an intermediate step of some larger distributed data processing pipeline. Such intermediate graph-structured data are already partitioned upon creation and, thus, distributed algorithms are essential in order to efficiently analyze them and avoid expensive data transfers.

The increasing interest in distributed graph processing is largely visible in both academia and industry. 
Research and industrial papers on parallel and distributed graph processing have seen a rise over the past few years~\cite{delft-tech-report,vertex-centric-survey} and dominate the proceedings of prime data management conferences~\cite{GiraphUC,Pregelix,Blogel,Giraph-VLDB,Graph-MR-SIGMOD,Graph-Maze-SIGMOD,KalavriVLDB,BigGraphsVLDB}. At the same time, open-source distributed graph processing systems are gaining popularity~\cite{Giraph-VLDB,Hama,GraphLab}, while general-purpose distributed data processing systems offer implementations of graph libraries and connectors to graph databases~\cite{GraphX,Gelly-post,Pregelix,Tinkerpop}.

Writing distributed graph mining applications is inherently hard. Computation parallelization, data partitioning, and communication management are major challenges of developing efficient distributed graph algorithms.
Furthermore, graph applications are highly diverse and expose a variety of data access and communication patterns~\cite{graph-processing-challenges,io-graph-algorithms}. For example, iterative refinement algorithms, like PageRank, can be expressed as parallel computations over the local neighborhood of each vertex, graph traversal algorithms produce unpredictable access patterns, while graph aggregations require grouping of similar vertices or edges together.

To address the challenges of distributed graph processing, several high-level programming abstractions and respective system implementations have been recently proposed~\cite{GraphLab,Pregel,Giraph-Plus,Arabesque,NScale}. Even though some have gained more popularity than others, each abstraction is optimized for certain classes of graph applications. For instance, the popular vertex-centric model~\cite{Pregel} is well-suitable for iterative value propagation algorithms, while the neighborhood-centric model~\cite{NScale} is designed to efficiently support operations on custom subgraphs, like ego networks. Unfortunately, there is no single model yet that can efficiently and intuitively cover all classes of graph applications.

Although several studies have experimentally compared the available distributed graph frameworks~\cite{benchmark-vision,Graphalytics,experimental-survey-springer,Study-Delft,experimental-study-vldb}, there exists no \emph{qualitative} comparison of the graph programming abstractions they offer. In this survey, we review prevalent high-level abstractions for distributed graph processing, in terms of semantics, expressiveness, and applicability. We identify the classes of graph applications that can be naturally expressed by each abstraction and we give examples of application domains for which a model appears to be non-intuitive. We further analyze popular distributed graph processing systems, with regards to their implementations of programming models and semantic restrictions. Our analysis can help graph application developers choose the appropriate model and system for their use-cases, and provides distributed systems researchers with open questions and challenges in the graph processing area.

This survey makes the following contributions.
\begin{itemize}
\item We present an overview of high-level programming abstractions for distributed graph processing and analyze their execution semantics and user-facing interfaces. We further consider performance limitations of graph programming models and we summarize proposed extensions and optimizations.
\item For each model, we identify classes of graph applications that can be intuitively expressed. Additionally, we look into examples of applications that are hard or problematic to express with each graph processing abstraction.
\item We categorize 34 distributed graph processing systems, with reference to the graph programming models they expose. We discuss execution models and communication mechanisms used.
\end{itemize}

The rest of this survey is organized as follows. In Section~\ref{sec:abstractions}, we describe several high-level programming abstractions for distributed graph processing. For each abstraction, we discuss semantics, applications, and existing performance optimizations. In Section~\ref{sec:categorization} we categorize distributed graph processing systems implementations, with regard to the programming model they expose. We discuss our findings, challenges and open questions in the area of distributed graph processing in Section~\ref{sec:open}. We review related work in Section~\ref{sec:related} and conclude this survey in Section~\ref{sec:conclusion}.

\subsection{Notation}
Here, we introduce the notation used for the execution semantics pseudocode and interfaces in the rest of this survey. Let $G=(V, E)$ be a directed graph, where $V$ denotes the set of vertices and $E$ denotes the set of edges. An edge is represented as a pair of vertices, where the first vertex is the source and the second vertex is the target. For example, for $u, v\in V$, an edge from $u$ to $v$ is represented by the pair $(u, v)$. Vertices and edges might have associated state (value) and local data structures of arbitrary type. For $v\in V$, $S_{v}$ refers to $v's$ associated value and $S_{(u,v)}$ refers to the value associated with the edge $(u,v)$. We define the set of first-hop, \emph{out-neighbors} of vertex $v$ as $N^{in}_{v} = \lbrace u | u \in V \land (v, u) \in E \rbrace$ and the set of first-hop \emph{in-neighbors} of $v$ as $N^{out}_{v} = \lbrace u | u \in V \land (u, v) \in E \rbrace$.

For the user-facing APIs, we use Java-like notation. In the rest of this survey, we assume that each vertex can be identified by a unique ID of an arbitrary type, to which we refer with $I$. Similarly, we use $VV$ to refer to the type of the vertex state, $EV$ for the type of the edge value, $M$ for message types, and $T$ for arbitrary intermediate data types.

\section{Programming Abstractions for Distributed Graph Processing}~\label{sec:abstractions}
In this section, we review high-level programming models for distributed graph processing. 
A distributed graph programming model is an abstraction of the underlying computing infrastructure that allows for the definition of graph data structures and the expression of graph algorithms.
We consider a distributed programming model to be \emph{high-level} if it hides data partitioning and communication mechanisms from the end user. Thus, programmers can concentrate on the logic of their algorithms and do not have to care about data representation, communication patterns, and underlying system architecture. High-level programming models are inevitably less flexible than low-level models, and limit the degree of customization they allow. On the other hand, they offer simplicity and facilitate the development of automatic optimization.

We provide a high-level description of main abstractions and user-facing APIs. We give examples of representative applications examples of algorithms that are difficult to express with certain abstractions. We also review implementation variants, known performance limitations, and proposed optimizations. Here, we describe six models that were developed \emph{specifically} for distributed graph processing, namely the vertex-centric, scatter-gather, gather-sum-apply-scatter, subgraph-centric, filter-process, and graph traversals.

\subsection{Vertex-Centric}
The vertex-centric model, introduced in the Pregel paper~\cite{Pregel}, is one of the most popular abstractions for large-scale distributed graph processing. Inspired by the simplicity of MapReduce~\cite{MapReduce}, which only requires the user to implement two functions, map and reduce, while hiding the complexity of communication and data distribution, the vertex-centric model places the vertex at the center of the computation. Also known as the \emph{think like a vertex}-model, it forces the user to express the computation from the point of view of a single vertex, by providing a single higher-order function.

A vertex-centric program receives a directed graph and a vertex function as input. A vertex serves as the unit of parallelization and has local state that consists of a unique ID, an optional vertex value, and its out-going edges, with optional edge values. Vertices communicate with other vertices through messages. A vertex can send a message to any other vertex in the graph, provided that it knows the destination's unique ID.

\begin{figure*}[!t]
\centering
\includegraphics[width=6.4in]{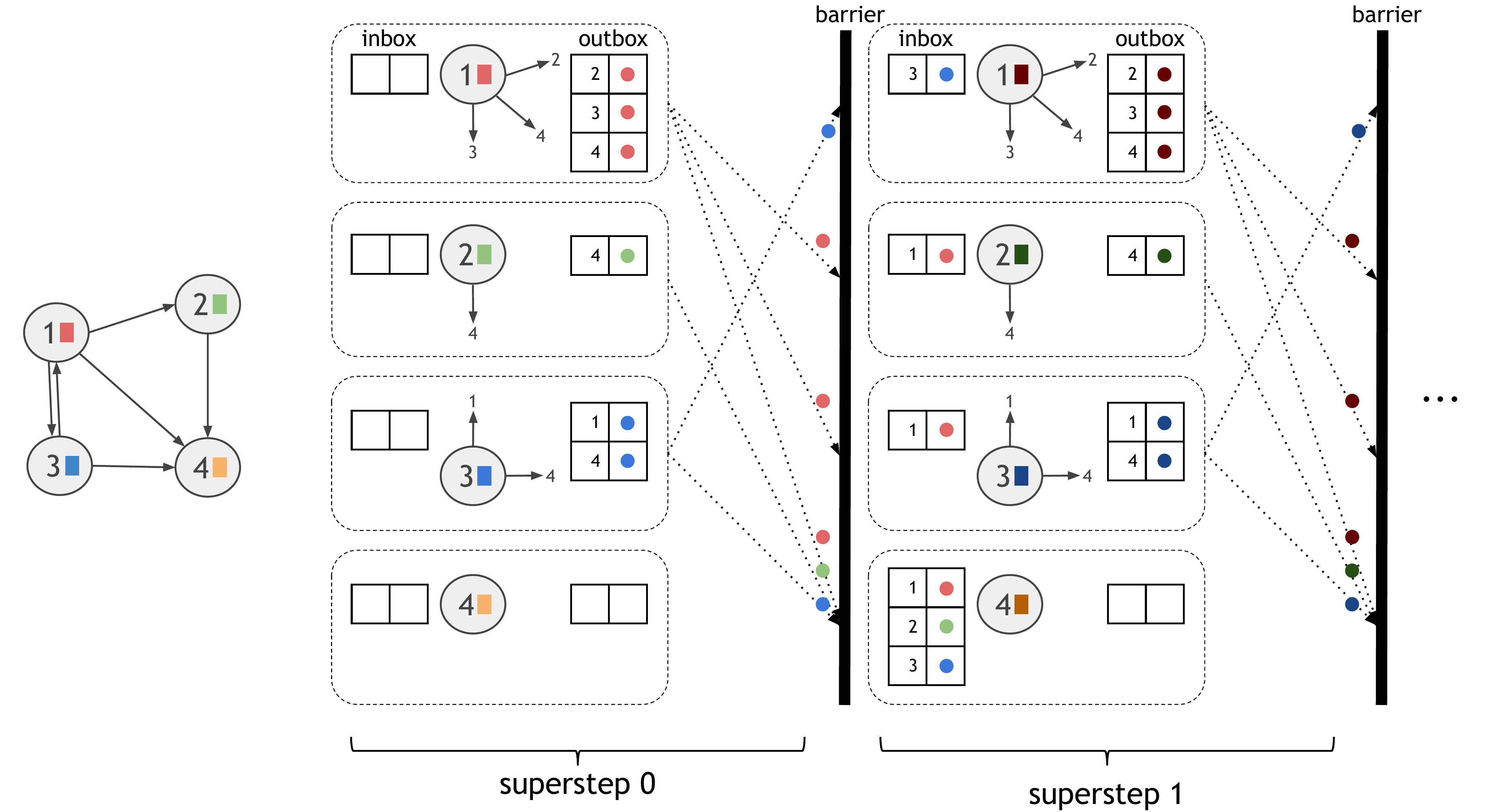}
\caption{Superstep execution and message-passing in the vertex-centric model. Graph edges can also have associated values, but we omit them here for simplicity. Each dotted box represents the computation scope of a vertex. Vertices in different scopes might reside on the same or different physical partitions. Arrows denote communication actions.}
\label{fig:vc-supersteps}
\end{figure*}

The execution workflow and computation parallelization of the vertex-centric model is shown in Figure~\ref{fig:vc-supersteps}. The dotted boxes correspond to parallelization units. 
Vertices can be in two states: \emph{active} or \emph{inactive}. Initially, all vertices are active. The computation proceeds in synchronized iterations, called \emph{supersteps}. In each superstep, all active vertices execute the same user-defined computation in parallel, called the \emph{vertex-function}. Supersteps are executed synchronously, so that messages sent during one superstep are guaranteed to be delivered in the beginning of the next superstep (be available to the vertex function of the vertex who is the receiver of the message). The output of a vertex-centric program is the set of vertex values at the end of the computation. If the graph is partitioned over several machines, a partition can contain several vertices and may have multiple worker-threads executing the vertex functions.

For simplicity of presentation, we define two auxiliary local data structures for each vertex. During superstep $i$, $inbox_{v}$ contains all the messages that were sent to vertex $v$ during superstep $i-1$. $outbox_{v}$ stores all the messages that vertex $v$ produces during superstep $i$. Message-passing is done as follows. At the end of each superstep, the runtime takes care of message delivery, by going through the $outbox$ of each sending vertex and placing the corresponding messages to the $inbox$ of each destination vertex. Message-passing can be implemented in batch or pipelined fashion. In the first case, messages are buffered in the local outbox of each vertex and are delivered in batches at the end of each superstep. In the case of pipelining, the runtime delivers messages to destination vertices as soon as they are produced. Pipelined message-passing might improve performance and lower memory requirements, but it limits ability of pre-aggregating the results at the outbox.

The execution semantics of the vertex-centric model are shown in Algorthm~\ref{algo:vc-semantics}. Initially, all vertices are active. At the beginning of each superstep, the runtime delivers messages to vertices ($receiveMessages$ function). Then, the user-defined vertex-function $compute$ is invoked in parallel for each vertex. It receives a set of messages as input and can produce one or more messages as output. At the end of a superstep, the runtime receives the messages from the $outbox$ of each vertex and computes the set of active vertices for the superstep. The execution terminates when there are no active vertices or when some user-defined convergence condition is met.

\begin{algorithm}
\caption{Vertex-Centric Model Semantics} \label{algo:vc-semantics}
\begin{algorithmic}
\State \textbf{Input:} directed graph G=(V, E)
\State $activeVertices\gets V$
\State $superstep\gets 0$
\While{$activeVertices\neq\emptyset$}
\For{$v\in activeVertices$}
\State $inbox_{v}\gets receiveMessages(v)$
\State $outbox_{v} = compute(inbox_{v})$
\EndFor
\State $superstep \gets superstep+1$
\EndWhile
\end{algorithmic}
\end{algorithm}

The user-facing interface of a vertex-centric program is shown in Interface 1.
The user-defined vertex function can read and update the vertex value and has access to all out-going edges. It can send a message to any vertex in the graph, addressing it by its unique ID. A vertex declares that it wants to become inactive by \emph{voting to halt}. If an inactive vertex receives a message, it becomes active again. In many implementations of the vertex-centric model, vertices can also add or remove a local edge or issue a mutation request for adding or removing non-local edges or vertices.

\begin{algorithm}
\floatname{algorithm}{Interface 1}
\renewcommand{\thealgorithm}{}
\caption{Vertex-Centric Model}
void abstract \textbf{compute}(Iterator[M] messages);

VV \textbf{getValue}();

void \textbf{setValue}(VV newValue);

void \textbf{sendMessageTo}(I target, M message);

Iterator \textbf{getOutEdges}();

int \textbf{superstep}();

void \textbf{voteToHalt}();
\end{algorithm}

\subsubsection{The GraphLab variant}
GraphLab~\cite{GraphLab} generalizes the vertex-centric model by introducing the notion of a vertex \emph{scope}. The scope of a vertex contains the adjacent edges, as well as the values of adjacent vertices. The vertex function is applied over the current state of a vertex and its scope. It returns updated values for the scope (a vertex can mutate the state of its neighbors) and a set of vertices \emph{T}, which will be scheduled for execution.
 
We observe the equivalence mapping between GraphLab's vertex-centric model and the Pregel, described in the previous section. In Pregel, the scope corresponds to the local vertex state together with the messages received from the neighboring vertices, and the set \emph{T} contains the active vertices. Instead of message-passing, GraphLab implements the \emph{pull} model, where vertices can read the values of neighbors in a defined \emph{scope}, and uses a shared-memory model to enable communication between vertices. GraphLab's vertex-centric programming model variant, allows for \textit{dynamic computation}, different consistency models and asynchronous execution. However, is also poses two limitations: (1) vertices can only communicate with their immediate neighbors, and (2) the graph structure has to be static, so that no mutations are allowed during execution. The shared-memory abstraction for communication is also adopted by Cyclops~\cite{cyclops}, even though its model is more restricted than the one provided by GraphLab.

\subsubsection{Applicability and expressiveness}
The vertex-centic model is general enough to express a broad set of graph algorithms.
The model is a good fit when the computation can be expressed as a local vertex function which only needs to access data on adjacent vertices and edges. Iterative value-propagation algorithms and fixed point methods map naturally to the vertex-centric abstraction.

A representative algorithm that can be easily expressed in the vertex-centric model is PageRank~\cite{PageRank}. In this algorithm, each vertex iteratively updates its rank by applying a formula on the sum of the ranks of its neighbors. The pseudocode for the PageRank vertex function is shown in Algorithm~\ref{algo:vc-pagerank}. Initially, all ranks are set to $1/numVertices()$. In each superstep, vertices send their partial rank along their outgoing edges and use the received partial ranks from their neighbors to update their ranks, according to the PageRank formula. After a certain number of supersteps (30 in this example) all vertices vote to halt and the algorithm terminates. In this pseudocode, we see that the superstep number is used to differentiate the computation between the first and the rest of the iterations. This is a common pattern in vertex-centric applications, since during superstep 0, no messages have been received by any vertex yet.

\begin{algorithm}
\caption{PageRank Vertex Function} \label{algo:vc-pagerank}
void \textbf{compute}(Iterator[double] messages):
\begin{algorithmic}
\State $outE=getOutEdges().size()$
\If{$superstep() > 0$}
\State \textbf{double} $sum=0$
\For{$m \in messages$}
\State $sum \gets sum + m$
\EndFor
\State $setValue(0.15/numVertices() + 0.85*sum)$
\EndIf
\If{$superstep() < 30$}
\For{$e \in getOutEdges()$}
\State $sendMessageTo(e.target(), getValue() / outE)$
\EndFor
\Else
\State $voteToHalt()$
\EndIf
\end{algorithmic}
\end{algorithm}

Non-iterative graph algorithms might be difficult to express in the vertex-centric model which heavily relies on the concept of supersteps. Furthermore, expressing a computation from the perspective of a vertex can often be challenging, as it also requires expressing all non-local state updates (further than one hop) as messages. Graph transformations and single-pass graph algorithms, like triangle counting, are not a good fit for the vertex-centric model.

\begin{figure}[!t]
\centering
\includegraphics[width=3.2in]{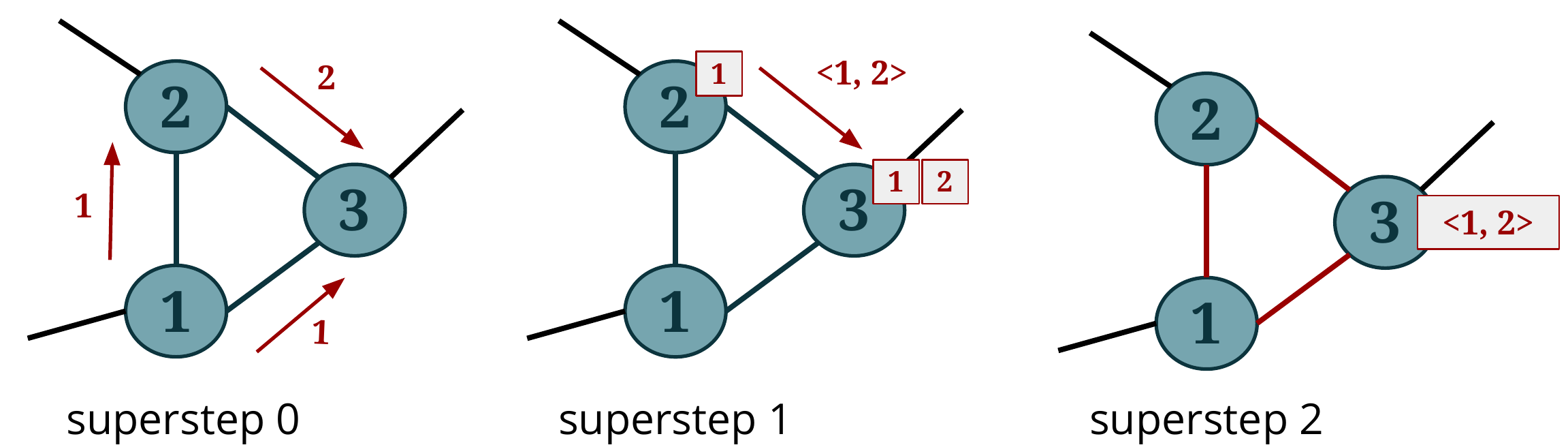}
\caption{A triangle counting algorithm in the vertex-centric model.
Messages produced by a vertex during the current superstep are shown with red arrows. Messages received in the current superstep are shown in grey boxes.}
\label{fig:vc-triangles}
\end{figure}

Let us consider an implementation of a triangle counting algorithm in the vertex-centric model. A triangle consists of three vertices which all form edges between them. A simple algorithm that counts all triangles of a graph in the vertex-centric model would be having each vertex count all the triangles where it appears. The main idea is to propagate a message along the edges of a triangle, so that when the message returns to the originator vertex, the triangle can be detected. In order not to count the same triangle multiple times, messages are only propagated from vertices with lower IDs to vertices with higher IDs. The algorithm is shown in Figure~\ref{fig:vc-triangles} and proceeds in three supersteps. During the first superstep, each vertex sends its ID to all neighbors with higher ID than its own. During the second superstep, each vertex attaches its own ID to every received message and propagates the pair of IDs to neighbors with higher IDs. During the final superstep, each vertex checks the received pairs of IDs to detect whether a triangle exists.

Another non-intuitive computation pattern is sending messages to the in-neighbors of a vertex; neighbors . Computing strongly connected components is an algorithm that contains this pattern~\cite{pregel-salihoglu}. Remember that each vertex only has access to its out-going edges and can only send messages to vertices with a known ID. Thus, if a vertex needs to communicate with its in-neighbors, it has to use a pre-processing superstep, during which, each vertex sends a message containing its own ID to all its out-neighbors. This way, all vertices will know the IDs of all their in-neighbors in the following superstep.

\subsubsection{Performance optimizations}
The wide spread of the vertex-centric modeldue to its simplicity and linear scalability for iterative graph algorithms, has spawned a large body of recent research and engineering efforts on optimizing its performance. In this section, we summarize some of the results of work on performance optimization for vertex-centric programs.

Communication can often become a bottleneck in the vertex-centric message-passing model. An overview of the model's limitations with regard to communication bottlenecks and worker load imbalance is presented in~\cite{pregel-plus}. The authors show that high-degree vertices or custom algorithm logic can create communication and computation skew, so that a small number of vertices produces many more messages than the rest, thus also increasing the workload of the worker machines where they reside. They tackle these problems with two message reduction techniques. First, they use \emph{mirroring}, a mechanism that creates copies of high-degree vertices on different machines, so that communication with neighbors can be local. A similar technique is also presented in~\cite{GPS}. Second, they introduce a \emph{request-response} mechanism, that allows a vertex to request the value of any other vertex in the graph, even if they are not neighbors. For not neighboring vertices, such a process would require three supersteps in the vanilla vertex-centric model. High communication load can also be avoided by using sophisticated partitioning mechanisms~\cite{partition-socc,GPS}.

Using synchronization barriers in the vertex-centric model allows programmers to write deterministic programs and easily reason about and debug their code. Even though using barriers for synchronization allows facilitating parallel programming for many applications, global barriers limit concurrency and may cause unnecessary over-synchronization, and as a consequence, poor performance, especially for applications with irregular or dynamic parallelism. In fact, various graph algorithms can benefit from asynchronous~\cite{sync-async,bertsekas,GraphLab} or hybrid~\cite{grace,spinning} execution models. In~\cite{GiraphUC}, the authors propose the \emph{barrierless asynchronous parallel} {BAP} model, to reduce stale messages and the frequency of global synchronization. The model allows vertices to immediately access messages they have received and utilizes only barriers local to each worker, which do not require global coordination.

Several algorithm-specific optimization techniques for the vertex-centric model are proposed in~\cite{pregel-salihoglu,Pregel-connectivity}. Specifically, the authors exploit the phenomenon of \emph{asymmetric convergence} often encountered in graph algorithms. We say that an iterative fixpoint graph algorithm converges asymmetrically, if different parts of the graph converge at different speeds. As a result, the overall algorithm converges slowly because, during the final supersteps,  only a small fraction of vertices are still active. The proposed solution is to monitor the active portion of the graph and, when it drops under some threshold, ship it to the master node, which executes the rest of the computation. Other optimizations include trading memory for communication and performing mutations (edge deletions) lazily.

In~\cite{Graph-Maze-SIGMOD}, the authors propose using special data structures, such as bit-vectors, to represent the neighborhood of each vertex. They also suggest compressing intermediate data that needs to be communicated over the network. They find that overlapping computation and communication, sophisticated partitioning, and different message-passing mechanisms have a significant impact on performance. Another issue they identify is that many algorithms have high memory requirements because of the outbox data structures of vertices growing too large. The authors propose to break each superstep into a number of smaller supersteps and processing only a subset of the vertices in each smaller superstep. Similar techniques are used in~\cite{Giraph-VLDB,KalavriVLDB}.

\subsection{Scatter-Gather}
The \emph{Scatter-Gather} abstraction, also known as \emph{Signal-Collect}~\cite{Signal-Collect}, is a vertex-parallel model, sharing the same \emph{think like a vertex} philosophy as the vertex-centric model. Scatter-Gather also operates in synchronized iteration steps and uses a message-passing mechanism for communication between vertices. The main difference is that each iteration step contains two computation phases, \emph{scatter} and \emph{gather}. Thus, the user also has to provide two higher-order functions, one for each phase.

\begin{figure}[!t]
\centering
\includegraphics[width=3.2in]{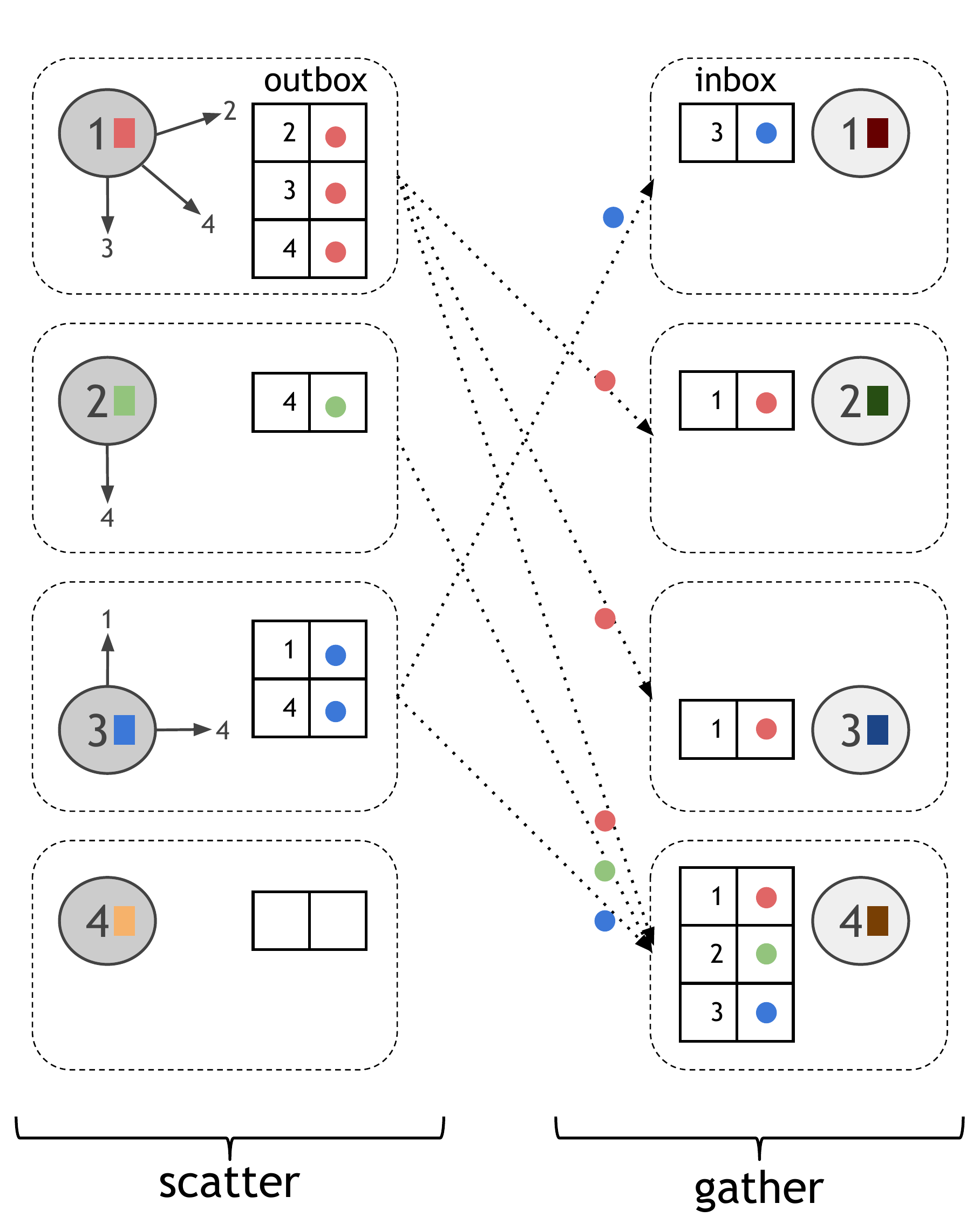}
\caption{One iteration in the Scatter-Gather model. In the Scatter phase, each vertex has read-access to its state, write-access to its outbox, and no access to its inbox. In the Gather phase, a vertex has read-access to its inbox, write-access to its state, but no access to its outbox.}
\label{fig:scatter-gather}
\end{figure}

The model is graphically shown in Figure~\ref{fig:scatter-gather}. Scatter-Gather decouples the sending of messages from the collection of messages and state update. During the scatter phase, each vertex executes a user-defined function that sends messages or \emph{signals} along out-going edges. During the gather phase, each vertex collects messages from neighbors and executes a second user-defined function that uses the received messages to update the vertex state. It is important to note that, contrary to the vertex-centric model, in Scatter-Gather both message sending and receipt happen during the \emph{same} iteration step. That is, during iteration $i$, the gather phase has access to the messages sent in the scatter phase of iteration $i$.

The execution semantics of the Scatter-Gather abstraction are shown in Algorthm~\ref{algo:sc-semantics}. The input of a Scatter-Gather program is a directed graph and the output is the state of the vertices after a maximum number of iterations or some custom convergence condition has been met. Similarly to the vertex-centric model, vertices can be in an active or inactive state. Initially, all vertices are active. Vertices can either explicitly vote to halt, like in the vertex-centric model, or implicitly get deactivated, if their value does not change during an iteration step. If a vertex does not update its value during a gather phase, then it does not have to execute a scatter phase in the next iteration, because its neighbors have already received its latest information. Thus, it gets deactivated.

\begin{algorithm}
\caption{Scatter-Gather Model Semantics} \label{algo:sc-semantics}
\begin{algorithmic}
\State \textbf{Input:} directed graph G=(V, E)
\State $activeVertices\gets V$
\State $superstep\gets 0$
\While{$activeVertices\neq \emptyset$}
\State $activeVertices^{'}\gets \emptyset$
\For{$v\in activeVertices$}
\State $outbox_{v} \gets scatter(v)$
\State $S_{v}^{'}\gets gather(inbox_{v}, S_{v})$
\If{$S_{v}^{'} \neq S_{v}$}
\State $S_{v}\gets S_{v}^{'}$
\State $activeVertices^{'} \gets activeVertices^{'} \cup {v}$
\EndIf
\EndFor
\State $activeVertices\gets activeVertices^{'}$
\State $superstep\gets superstep+1$
\EndWhile
\end{algorithmic}
\end{algorithm}

The user-facing interfaces of Scatter and Gather are shown in Interfaces 2 and 3, respectively. Note that the scopes of the two phases are separate and each interface has different available methods.
The scatter interface can retrieve the current vertex value, read the state of the neighboring edges, and send messages to neighboring vertices. The gather interface can access received messages, read and set the vertex value.

\begin{algorithm}
\floatname{algorithm}{Interface 2}
\renewcommand{\thealgorithm}{}
\caption{Scatter}
void abstract \textbf{scatter}();

VV \textbf{getValue}();

void \textbf{sendMessageTo}(I target, M message);

Iterator \textbf{getOutEdges}();

int \textbf{superstep}();
\end{algorithm}

\begin{algorithm}
\floatname{algorithm}{Interface 3}
\renewcommand{\thealgorithm}{}
\caption{Gather}
void abstract \textbf{gather}(Iterator[M] messages);

void \textbf{setValue}(VV newValue);

VV \textbf{getValue}();

int \textbf{superstep}();
\end{algorithm}

\subsubsection{Applicability and expressiveness}
Scatter-Gather is a useful abstraction and can be used to express a variety of algorithms in a concise and elegant way. Similarly to the vertex-centric model, iterative, value-propagation algorithms like PageRank are a good fit for Scatter-Gather. Since the logic of producing messages is decoupled from the logic of updating vertex values based on the received messages, programs written using Scatter-Gather are sometimes easier to follow and maintain. The vertex-centric PageRank example that we saw in the previous section can be easily expressed in the Scatter-Gather model, by simply separating the sending of messages and the calculation of ranks in two phases. The pseudocode is shown in Algorithm~\ref{algo:sg-pagerank}. The scatter phase contains only the the rank propagation logic, while the gather phase contains the message processing and rank update logic.

\begin{algorithm}
\caption{PageRank Scatter and Gather Functions} \label{algo:sg-pagerank}
void \textbf{scatter}():
\begin{algorithmic}
\State $outEdges=getOutEdges().size()$
\For{$edge \in getOutEdges()$}
\State $sendMessageTo(edge.target(), getValue() / outEdges)$
\EndFor
\end{algorithmic}
void \textbf{gather}(Iterator[\textbf{double}] messages):
\begin{algorithmic}
\If{$superstep() < 30$}
\State \textbf{double} $sum=0$
\For{$m \in messages$}
\State $sum \gets sum + m$
\EndFor
\State $setValue(0.15/numVertices() + 0.85*sum)$
\EndIf
\end{algorithmic}
\end{algorithm}

Separating the messaging phase from the vertex value update logic not only makes some programs easier to follow but might also have a positive impact on performance. Scatter-Gather implementations typically have lower memory requirements, because concurrent access to the inbox (messages received) and outbox (messages to send) data structures is not required. However, this characteristic also limits expressiveness and makes some computation patterns non-intuitive. If an algorithm requires a vertex to concurrently access its inbox and outbox, then the expression of this algorithm in Scatter-Gather might be problematic. Strongly Connected Components and Approximate Maximum Weight Matching~\cite{pregel-salihoglu} are examples of such graph algorithms. A direct consequence of this restriction is that vertices cannot generate messages and update their states in the same phase. In the scatter phase, vertices have read-access to their state and adjacent edges, write-access to their outbox, and no access to their inbox. In the gather phase, vertices have read-access to their inbox, write-access to their state, but no access to their outbox or adjacent edges. Thus, deciding whether to propagate a message based on its content would require storing it in the vertex value, so that the gather phase has access to it, in the following iteration step. Similarly, if the vertex update logic includes computation over the values of the neighboring edges, these have to be included inside a special message passed from the scatter to the gather phase. Such workarounds often lead to higher memory requirements and non-elegant, hard to understand algorithm implementations.

For example, consider the problem of finding whether there exists a path from source vertex $a$ to target vertex $b$, with total distance equal to a specified user value $d$. Let us assume that the input graph has edges with positive values corresponding to distances. We can solve this problem in a message-passing vertex-parallel way, by iteratively propagating messages through the graph and aggregating edge weights on the way. Messages originate from the source vertex and are routed towards the target vertex, one neighborhood hop per superstep. When vertex $v$ receives a message, it decides whether to propagate the message to a neighbor $u$, based on the current distance that the message contains plus the distance of the edge that connects $v$ with $u$. If the computed sum is less than or equal to $d$, $v$ updates the message content and propagates it to $u$. If the sum exceeds the value of $d$, then $v$ drops the message. In the vertex-centric model, this logic can be implemented inside the vertex compute function, since vertices receive and send messages during the same phase. However, in Gather-Scatter, vertices receive messages in the gather phase, but can only generate messages in the scatter phase. In order for a vertex to know whether to propagate a message based on its content, it needs a mechanism to allow the scatter phase access messages received in the previous superstep. One way that this can be achieved is by storing all received messages in the vertex value, so that the scatter interface can access them in the next superstep.

\subsection{Gather-Sum-Apply-Scatter (GAS)}
The Gather-Sum-Apply-Scatter programming abstraction (GAS), first introduced by Powergraph~\cite{PowerGraph}, tries to address performance issues that arise when using the vertex-centric or scatter-gather model on power-law graphs. In such graphs, most vertices have relatively few neighbors, while few vertices have a very large number of neighbors. This degree \emph{skew} causes computational imbalance in vertex-parallel programming models. The few high-degree vertices, having much more work to do during a superstep, act as stragglers, thus, slowing down the overall execution.

\begin{figure*}[!t]
\centering
\includegraphics[width=4.8in]{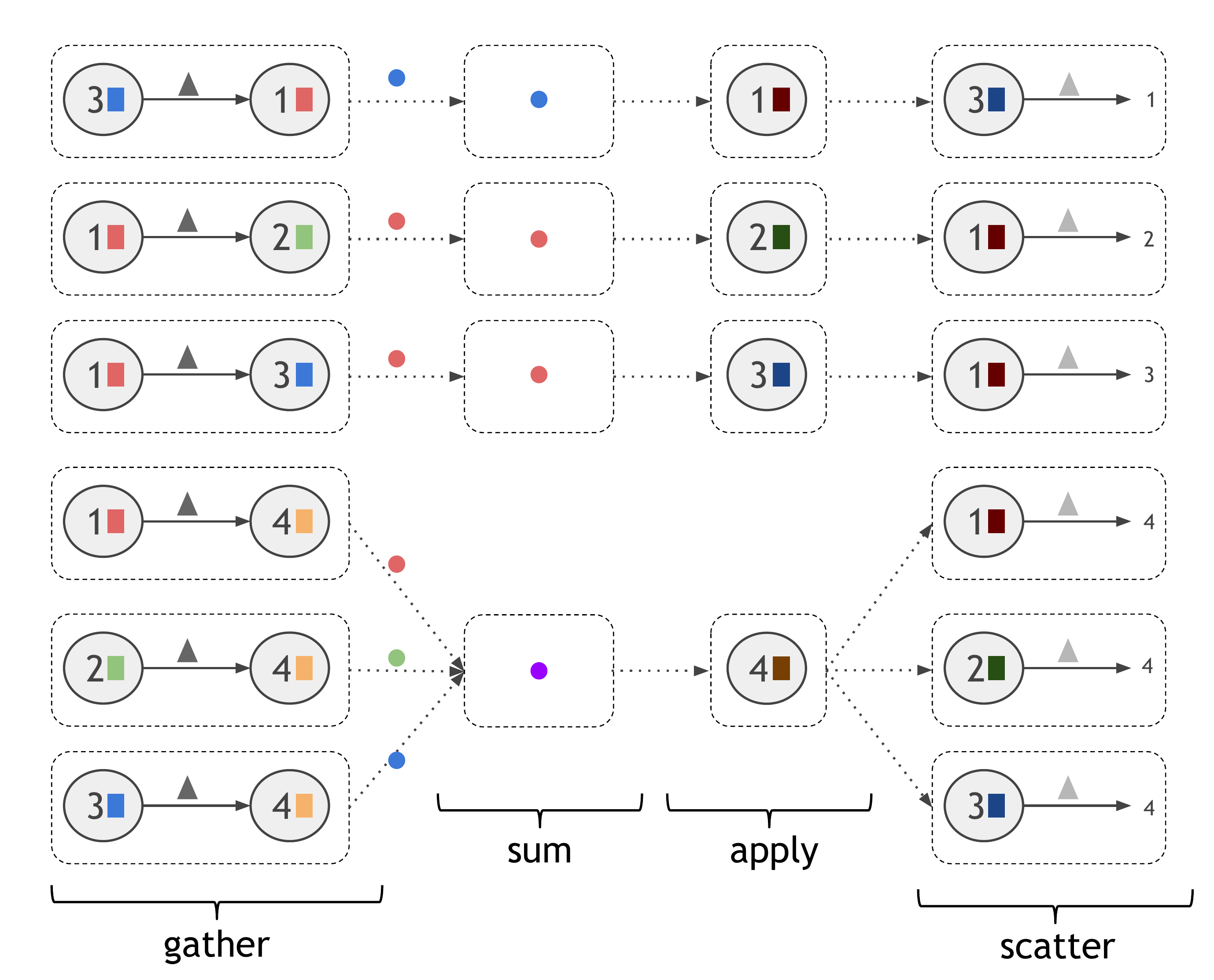}
\caption{The GAS computation phases. The gather phase parallelizes the computation over the edges of the input graph. The user-defined function has access to an edge, including its source and target vertex states. The sum phase combines partial values using a user-defined associative and commutative function. The vertex states are updated during the apply phase. The scatter phase is optional and can be used to update edge states.}
\label{fig:gsas}
\end{figure*}

The GAS model addresses the bottlenecks caused by high-degree vertices by parallelizing the computation over the \emph{edges} of the graph. The abstraction essentially decomposes a vertex-program in separate phases, which allow distributing the computation more effectively over a cluster. The computation proceeds in four phases, each executing a user-defined function. During the \emph{gather} phase, a user-defined function is applied on each of the adjacent edges of each vertex in parallel, where an edge contains both the source vertex and the target vertex values. The transformed edges are passed to an \emph{associative and commutative} user-defined function, which combines them to a single value during the \emph{sum} phase. The gather and sum phases naturally correspond to a \emph{map-reduce} step and are sometimes considered as a single phase~\cite{PowerGraph}. The result of the sum phase and the current state of each vertex are passed to the \emph{apply} user-defined function, which uses them to compute the new vertex state. During the final \emph{scatter} phase, a user-defined function is invoked in parallel per edge, having to access the updated source and target vertex values. In some implementations of the model, the scatter phase is either optional or omitted. The four phases are graphically shown in Figure~\ref{fig:gsas} and its execution semantics can be found in Algorithm~\ref{algo:gsas-semantics}.

\begin{algorithm}
\caption{GAS Model Semantics} \label{algo:gsas-semantics}
\begin{algorithmic}
\State \textbf{Input:} directed graph G=(V, E)
\State $a_{v}\gets$ \textbf{empty}
\For{$v\in V$}
\For{$n\in N^{in}_{v}$}
\State $a_{v}\gets sum(a_{v}, gather(S_{v}, S_{(v, n)}, S_{n}))$
\EndFor
\State $S_{v}\gets apply(S_{v}, a_{v})$
\State $S_{(v, n)}\gets scatter(S_{v}, S_{(v, n)}, S_{n})$
\EndFor
\end{algorithmic}
\end{algorithm}

The public interfaces of the GAS abstraction are shown in Interface 4. Note how these interfaces are simpler and more restrictive than the vertex-centric and gather-scatter interfaces. All four user-defined functions return a single value and the scope of computation is restricted to local neighborhoods.

\begin{algorithm}
\floatname{algorithm}{Interface 4}
\renewcommand{\thealgorithm}{}
\caption{Gather-Sum-Apply-Scatter}
T abstract \textbf{gather}(VV sourceV, EV edgeV, VV targetV);

T abstract \textbf{sum}(T left, T right);

VV abstract \textbf{apply}(VV value, T sum);

EV abstract \textbf{scatter}(VV newV, EV edgeV, VV oldV);
\end{algorithm}

\subsubsection{Applicability and expressiveness}
The GAS abstraction imposes the restriction of an associative and commutative sum function to produce edge-parallel programs that will not suffer from computational skew. Nevertheless, the model
can be used to emulate vertex-centric programs, even if the update function is not associative and commutative. To express a vertex-centric computation, the gather and sum functions can be used to combine the inbound messages (stored as edge data) and concatenate the list of neighbors needed to compute the outbound messages. The vertex compute function is then executed inside the apply phase. The apply user-defined function generates the messages, which can then be passed as vertex data to the scatter phase. Similarly, to emulate a GraphLab vertex program, the gather and sum functions can be used to concatenate all the data on adjacent vertices and edges and then run the vertex function inside the apply phase. 

Executing vertex-parallel programs inside the apply function results in complexity linear in the vertex degree, thus defeating the purpose of eliminating computational skew. Moreover, manually constructing the neighborhood in the sum phase and concatenating messages in order to access them in the apply phase, are both non-intuitive and computationally expensive. Fortunately, many graph algorithms can be decomposed into a gather transformation and an associative and commutative sum function. Algorithm~\ref{algo:gsas-pagerank} shows a PageRank implementation using the GAS interfaces. The gather phase computes a partial rank for each neighbor. The sum phase sums up all the partial ranks into a single value, and the the apply phase computes the new PageRank and updates the vertex value. The scatter phase has been omitted, since edge values do not get updated in this algorithm. Alternatively, the scatter phase can be used to selectively activate vertices for the next iteration~\cite{PowerGraph}.

\begin{algorithm}
\caption{PageRank Gather, Sum, Apply} \label{algo:gsas-pagerank}
\textbf{double gather}(\textbf{double} src, \textbf{double} edge, \textbf{double} trg):
\begin{algorithmic}
\State \textbf{return} $trg.value / trg.outNeighbors$
\end{algorithmic}
\textbf{double} \textbf{sum}(\textbf{double} rank1, \textbf{double} rank2):
\begin{algorithmic}
\State \textbf{return} $rank1 + rank2$
\end{algorithmic}
\textbf{double} \textbf{apply}(\textbf{double} sum, \textbf{double} currentRank):
\begin{algorithmic}
\State \textbf{return} $0.15 + 0.85 * sum$
\end{algorithmic}
\end{algorithm}

If the algorithm cannot be decomposed into a gather step and a commutative-associative sum, implementation in the GAS model might require manual emulation of the vertex-centric or scatter-gather models, as described previously. For example, in the Label Propagation algorithm~\cite{LabelPropagation}, a vertex receives labels from its neighbors and chooses the most frequent label as its vertex value. Computing the most frequent item is not an associative-commutative function. In order to express this algorithm in GAS, the sum phase needs to construct a set of all the  neighbor labels. The gather user-defined function returns a set containing a single label for each neighbor. The sum user-defined function receives a pair of sets, each containing a neighbor's label and returns the sets' union. Finally, each vertex chooses the most frequent label in the apply function, which has access to all labels.

Table~\ref{tbl:vertex-scope-comparison} shows a comparison among the vertex-centric, gather-scatter, and GAS programming models, with regard to update functions and communication. We notice that the vertex-centric model is the most generic of the three, allowing for arbitrary vertex-update functions and communication logic, while GAS is the most restricted.

\begin{table*}[]
\centering
\caption{Comparison of the vertex-centric, gather-scatter, and GAS programming models.}
\label{tbl:vertex-scope-comparison}
\begin{tabular}{p{1.3cm}|p{2.6cm}|p{2.6cm}|p{2.4cm}|p{2.4cm}|}
\cline{2-5}
&  \textbf{update function properties} & \textbf{update function logic} &  \textbf{communication scope} &  \textbf{communication logic} \\ \hline \hline
\multicolumn{1}{|l|}{\textbf{Vertex-Centric}}                        &  arbitrary                           & arbitrary                      &  any vertex                   &  arbitrary                   \\ \hline
\multicolumn{1}{|l|}{ \textbf{Scatter-Gather}}                        &  arbitrary                           & based on received messages     &  any vertex                   &  based on vertex state       \\ \hline
\multicolumn{1}{|l|}{ \textbf{GAS}}           & associative and commutative             & based on neighbors' values     &  neighborhood                 & based on vertex state        \\ \hline
\end{tabular}
\end{table*}

\subsubsection{Performance optimizations}
In~\cite{PowerLyra}, the authors find that, even though the GAS model manages to overcome the load imbalance issues caused by high-degree vertices, at the same time it poses a high memory and communication overhead for the low-degree vertices of the input graph. They propose a \emph{differentiated vertex computation} model, where the high-degree vertices are processed using the GAS model, while the low-degree vertices are processed using a GraphLab-like vertex-centric model.

\subsection{Subgraph-Centric}
All the models we have seen so far, vertex-centric, gather-scatter, and GAS, operate on the scope of a single vertex or edge in the graph. While such fine-grained abstractions for distributed programming are considerably easy to use by non-experts, they might cause high communication overhead when compared to coarse-grained abstractions. In this section, we present two distributed graph processing models that operate on the \emph{subgraph} level, with the objective to reduce communication and exploit the subgraph structure to allow the implementation of optimized algorithms.

\subsubsection{Partition-Centric}
The partition-centric model offers a lower-level abstraction than the vertex-centric, setting the whole partition as the unit of parallel computation. The model exposes the subgraph of each partition to the user-function, in order to avoid redundant communication and accelerate convergence of vertex-centric programs. The abstraction has been recently introduced by Giraph++\cite{Giraph-Plus} and has been quickly adopted and further optimized in several succeeding works~\cite{GoFFish,Blogel}.

The main idea behind the partition-centric model relies on the perception of each partition as a proper subgraph of the input graph, instead of a collection of unassociated vertices. While in the vertex-centric model a vertex is restricted to accessing information from its immediate neighbors, in the partition-centric model information can be propagated freely inside all the vertices of the same partition. This simple property of the partition-centric model can lead to significant communication savings and faster convergence.

\begin{figure*}[!t]
\centering
\includegraphics[width=5.2in]{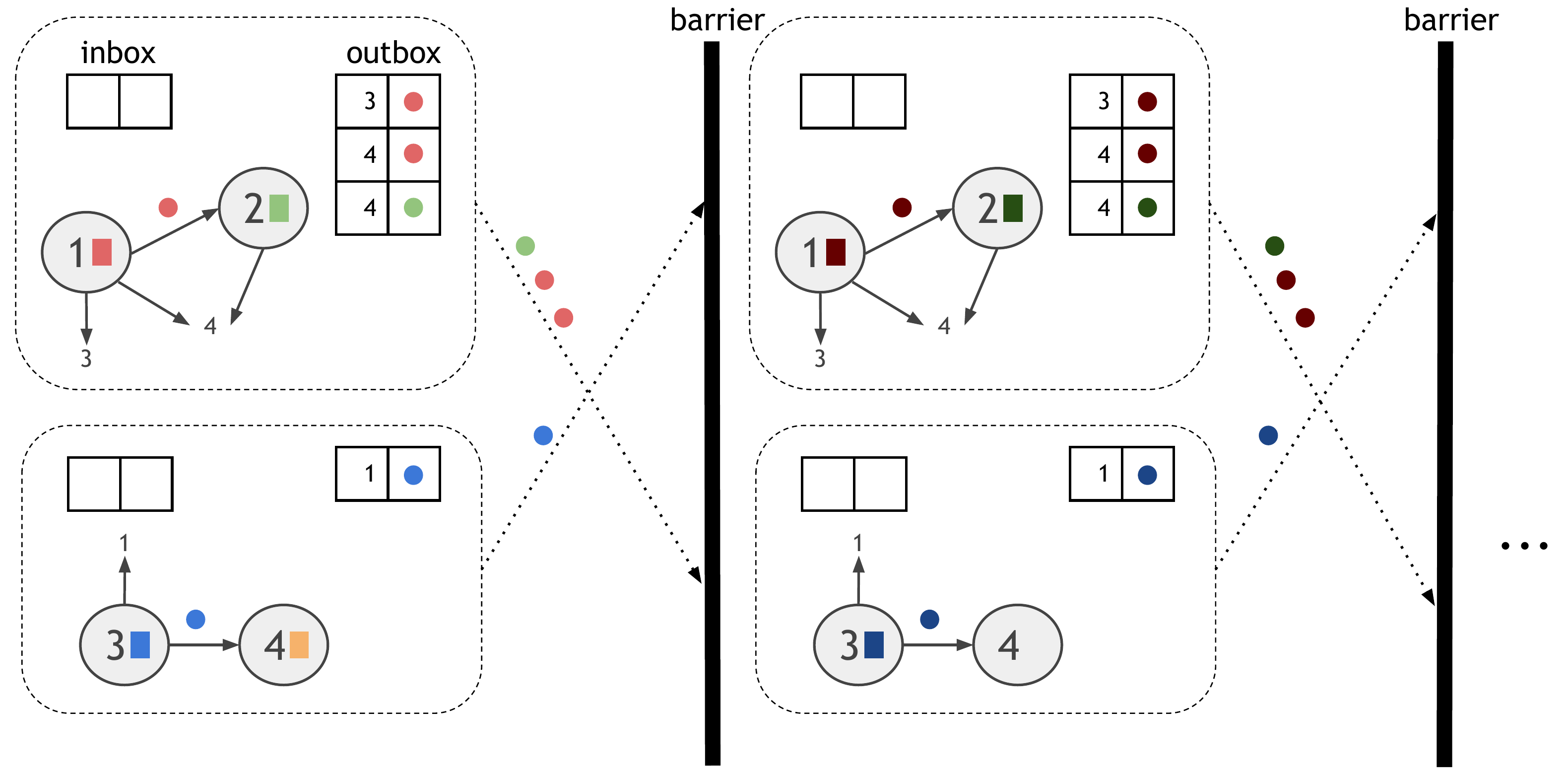}
\caption{Two iterations in the Partition-Centric model. Each dotted box represents a separate physical partition and dotted arrows represent communication.}
\label{fig:partition-centric}
\end{figure*}

The partition-centric model is graphically shown in Figure~\ref{fig:partition-centric}. Note that the whole partition becomes the parallelization unit on which the user-defined function is applied. As compared to the vertex-centric case, here, message exchange happens only between partitions, thus resulting in reduced communication costs. Inside a partition, vertices can be \emph{internal} or \emph{boundary}. Internal vertices are associated with their value, neighboring edges, and incoming messages. Boundary vertices only have a local \emph{copy} of their associated value; the primary value resides in the partition where the vertex is internal. In Figure~\ref{fig:partition-centric}, vertices $1$ and $2$ are internal in the upper partition, while vertices $3$ and $4$ are boundary. In the lower partition, vertices $3$ and $4$ are internal, while vertex $1$ is boundary. Message exchange between internal vertices of the same partition is immediate, while messages to boundary vertices require network transfer.

The execution semantics of the partition-centric model are the same as in the vertex-centric model, with the only difference that the user-defined update function is invoked per partition and messages are distributed between partitions, not vertices. The user-facing interface of the model offers most of the methods of the vertex-centric interface, with a few additional ones, shown below. The interface needs to provide a mechanism to check whether a vertex belongs to a particular partition and to retrieve internal and boundary vertices inside a partition. Note that since the scope of the user-defined function is not a single vertex, the methods to retrieve and update vertex attributes are not part of the \emph{compute()} interface.

\begin{algorithm}
\floatname{algorithm}{Interface 5}
\renewcommand{\thealgorithm}{}
\caption{Partition-Centric Model}
void abstract \textbf{compute}();

void \textbf{sendMessageTo}(I target, M message);

int \textbf{superstep}();

void \textbf{voteToHalt}();

boolean \textbf{containsVertex}(I id);

boolean \textbf{isInternalVertex}(I id);

boolean \textbf{isBoundaryVertex}(I id);

Collection \textbf{getInternalVertices}();

Collection \textbf{getBoundaryVertices}();

Collection \textbf{getAllVertices}();
\end{algorithm}

\subsubsection{Neighborhood-Centric}
The neighborhood-centric model~\cite{NScale} sets the scope of computation on \emph{custom} subgraphs of the input graph. These subgraphs are explicitly built around vertices and their multi-hop neighborhoods, in order to facilitate the implementation of graph algorithms that operate on ego-networks; networks built around a central vertex of interest. The user specifies custom subgraphs and a program to be executed on those subgraphs. The user program might be iterative and is executed in parallel on each subgraph, following the Bulk Synchronous protocol (BSP)~\cite{BSP}. In contrast to the partition-centric model, in an implementation of the neighborhood-centric model, a physical partition can contain one or more custom subgraphs.
Information exchange happens through shared state updates for subgraphs in the same partition and through replicas and messages for subgraphs belonging to different partitions.

\begin{figure}[!t]
\centering
\includegraphics[width=2.6in]{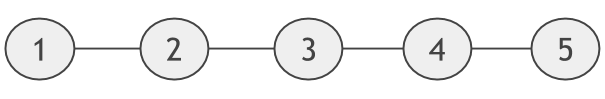}
\caption{A chain example graph.}
\label{fig:chain-graph}
\end{figure}

\begin{figure}[!t]
\centering
\includegraphics[width=3.2in]{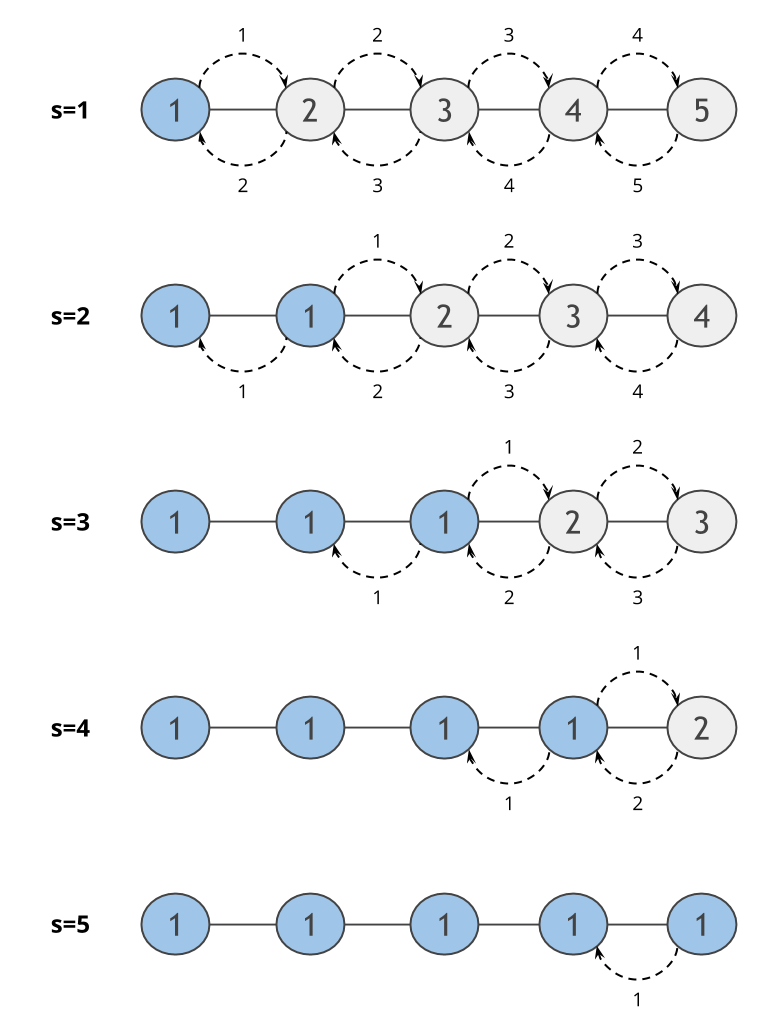}
\caption{Connected Components via label propagation in the vertex-centric model. The minimum value propagates to the end of the chain after 5 supersteps.}
\label{fig:chain-graph-vc}
\end{figure}

\begin{figure}[!t]
\centering
\includegraphics[width=3.2in]{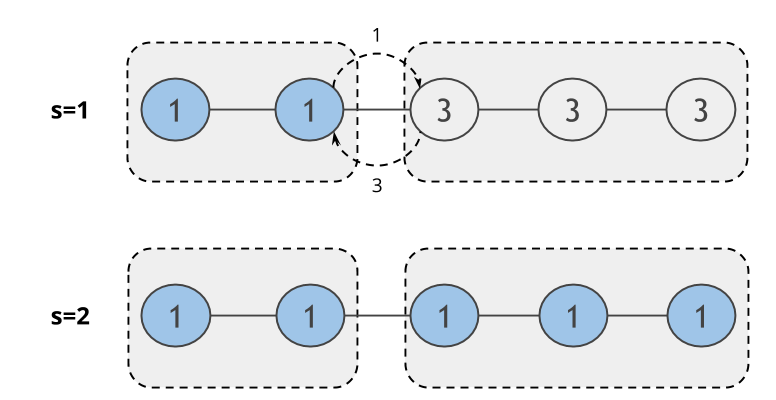}
\caption{Connected Components via label propagation in the partition-centric model. Vertices 1 and 2 belong to the first partition and vertices 3, 4, 5 belong to the second partition. Each partition converges asynchronously, before initiating communication with other partitions. The minimum value propagates to all vertices after 2 supersteps.}
\label{fig:chain-graph-pc}
\end{figure}

\subsubsection{Applicability and expressiveness}
The partition-centric model is a good fit when there exists an efficient sequential algorithm which can be executed in each partition and whose partial results can be easily combined to produce the global result. The performance of this model is highly dependent on the quality of the partitions. If the partitioning technique used creates well-connected subgraphs and minimizes edge cuts between partitions, it is highly probable that a partition-centric implementation will require less communication than a vertex-centric implementation and that a value-propagation algorithm will converge is fewer supersteps. For example, consider the execution of the connected components algorithm on a chain graph, like the one shown in Figure~\ref{fig:chain-graph}. In a vertex-centric execution of the algorithm, shown in Figure~\ref{fig:chain-graph-vc}, the minimum value can only propagate one hop per superstep. Consequently, the algorithm requires as many supersteps as the maximum graph diameter plus one, in order to converge. In the parition-centric model instead, values can propagate asynchronously inside a partition. If the chain is partitioned in two connected subgraphs, like the ones of Figure~\ref{fig:chain-graph-pc}, the algorithm converges after just two supersteps. However, if the graph is partitioned poorly, there will be little to no benefit compared to the vertex-centric execution. For instance, if we partition the given chain into two partitions of odd and even vertices, the algorithm will need as many supersteps to converge as in the vertex-centric case. Sophisticated partitioning can be an expensive task and users must carefully consider the pre-processing cost that it might impose to the total job execution time.

A drawback of the partition-centric model is that users have to switch from "thinking like a vertex" to thinking in terms of partitions. In order to reason about their algorithm, users need to understand what a partition represents and how to differentiate the behavior of an internal vertex versus that of a boundary vertex. The model allows for more control on computation and communication, but at the same time exposes low-level characteristics to users. This loss of abstraction might lead to erroneous or hard-to-understand programs.

Algorithm~\ref{algo:pc-pagerank} shows the pseudocode of a partition-centric PageRank implementation. It is immediately apparent that this implementation is quite longer and more complex than the ones we have seen so far. Part of the complexity is introduced from the fact that, inside each partition, PageRank computation is asynchronous. Each vertex has an additional attribute, \emph{delta}, besides its PageRank score, where it stores intermediate updates from vertices inside the same partition. At the end of each superstep, boundary vertices with positive \emph{delta} values produce messages to be delivered in other partitions.

\begin{algorithm}
\caption{PageRank Partition-Centric Function} \label{algo:pc-pagerank}
void \textbf{compute}():
\begin{algorithmic}
\If{$superstep() > 0$}
  \For{$v \in getAllVertices()$}
    \State $v.getValue().pr = 0$
    \State $v.getValue().delta = 0$
  \EndFor
\EndIf
\For{$iv \in internalVertices()$}
  \State $outEdges = iv.getNumOutEdges()$ 
  \If{$superstep() > 0$}
    \State $iv.getValue().delta += 0.15$
  \EndIf
  \State $iv.getValue().delta += iv.getMessages()$
  \If{$iv.getValue().delta > 0$}
    \State $iv.getValue().pr += iv.getValue().delta$
    \State $u=0.85*iv.getValue().delta/outEdges$
    \While{$iv.iterator.hasNext()$}
      \State $neighbor = getVertex(iv.iterator().next())$
      \State $neighbor.getValue().delta+=u$
    \EndWhile
  \EndIf
  \State $iv.getValue().delta=0$
\EndFor
\For{$bv \in boundaryVertices()$}
  \State $bvID = bv.getVertexId()$
  \If{$bv.getValue().delta > 0$}
    \State $sendMessageTo(bvID, bv.getValue().delta)$
    \State $bv.getValue().delta=0$
  \EndIf
\EndFor
\end{algorithmic}
\end{algorithm}

The neighborhood-centric model is preferable when applications require accessing multi-hop neighborhoods or ego-centric networks of certain vertices. Such applications include personalized recommendations, social circle analysis, anomaly detection, and link prediction. While in the partition-centric model, the graph is partitioned into non-overlapping, application-independent subgraphs, in the neighborhood-centric model, subgraphs are extracted based on specific application criteria. This way, users can specify subgraphs to be $k$-hop neighborhoods around a set of query vertices, satisfying a particular predicate. The user program executed on these subgraphs can have arbitrary random access to the state of the whole subgraph. Note that creating these overlapping neighborhoods requires an often expensive pre-processing step, in terms of both execution time and memory.

\subsubsection{Performance optimizations}
GoFFish~\cite{GoFFish} and Blogel~\cite{Blogel} propose a subgraph-centric model that partitions the input graph in subgraphs which are \emph{connected}. This way, well-known shared-memory algorithms can be applied on the connected subgraphs. A partition can contain multiple subgraphs and execute the computation on each subgraph in parallel. GoFFish also proposes a distributed persistent storage, optimized for subgraph access patterns. On the other hand, Blogel allows a subgraph to define and manage its own state, allowing for subgraph-level communication. Both GoFFish and Blogel are beneficial when the number of subgraphs is sufficiently larger than the number of workers, so that a balanced workload can be achieved.

\subsection{Filter-Process}
The \emph{filter-process} computational model, also known as or "think like an \emph{embedding}", is proposed by Arabesque~\cite{Arabesque}. Arabesque is a system for efficient distributed graph mining. An embedding is a subgraph instance of the input graph that matches a user-specified pattern. The model facilitates the development of graph mining algorithms, which require subgraph enumeration and exploration. Such algorithms are challenging to express and efficiently support with a vertex-centric model, due to immense intermediate state and high computation requirements.

The programming model consists of two primary functions, \emph{filter} and \emph{process}. \emph{Filter} examines whether a given embedding is eligible for processing and \emph{process} executes some action on the embedding and may produce output. The model assumes immutable input graphs and connected graph patterns.

Computation proceeds in a sequence of \emph{exploration steps}, using the BSP model. During each exploration step, the system explores and extends an input set of embeddings. First, a set of candidate embeddings is created, by extending the set of input embeddings. In the first exploration step the set of candidates contains all the edges or vertices of the input graph. Once the candidates have been produced, the filter function examines them and selects the ones that should be processed. The selected embeddings are then sent to the process function, which outputs user-defined values. Finally, the selected embeddings become the input set of embeddings for the next step. The computation terminates when there are no embeddings to be extended.

The filter-process model differs from the partition-centric and neighborhood-centric models, where partitions and subgraphs are generated \emph{once}, at the beginning of the computation as a pre-processing step. In filter-process, embeddings are dynamically generated during the execution of exploration steps. The model is suitable for graph pattern mining problems, which require subgraph enumeration, such as network motif discovery, semantic data processing, and spam detection.

\subsection{Graph Traversals}
Distributed graph analysis through \emph{graph traversals} is the programming model adopted by the Apache Tinkerpop project~\cite{Tinkerpop}. The system provides a graph traversal machine and language, called \emph{Gremlin}~\cite{gremlin}, which supports distributed traversals via the Bulk Synchronous Parallel (BSP) computation model.

In the traversal model of graph databases, \emph{traversers} walk through an input graph, following user-provided instructions. The Gremlin machine supports distributed graph traversals, by modeling traversers as messages. Vertices receive traversers, execute their traversal step, and, as a result, generate other traversers to be sent as messages to other vertices. Halted traversers are stored in a vertex attribute. The process terminates when no more traversers are being sent. The result of the computation is the aggregate of the locations of the halted traversers.

\section{General-Purpose Programming Models used for Graph Processing}
Except from the specialized programming models that we have reviewed so far, several general-purpose distributed programming models have also been used for graph processing. Here, we review five such data processing abstractions that have been used for developing graph applications, namely MapReduce, dataflow, linear algebra primitives, datalog, and shared partitioned tables.	

\subsection{MapReduce}
MapReduce~\cite{MapReduce} is a popular distributed programming model for large-scale data processing on commodity clusters. Inspired by functional programming, it provides two operators, \emph{map} and \emph{reduce}, which encapsulate user-defined functions and form a static pipeline. A MapReduce application typically reads input in the form of \emph{key-value} pairs from a distributed file system. Input pairs are processed by parallel map tasks, which apply the user-defined function, producing
intermediate key-value pairs. The intermediate results are sorted and grouped by key, so that each group is then processed by parallel reduce tasks, applying the reduce user-defined function. Pairs sharing the same key are sent to the same reduce task. The output of each reduce task is eventually written to the distributed file system, producing the job output.

MapReduce was quickly gained popularity because of its simplicity and scalability. However, its programming model is not suitable for graph applications, which are often iterative and require multi-step computations. Several extensions of the model have been proposed in order to support such algorithms~\cite{HaLoop,Twister,pegasus}.
The main idea of these extensions is adding a driver program that can coordinate iterations, containing one or more MapReduce jobs. The main task of the driver is to submit a new job per iteration and track convergence. Iterations are typically chained by using the output of one MapReduce-iteration as the input of the next.

Pegasus~\cite{pegasus} implements a generalized iterative matrix-vector multiplication primitive, GIM-V, as a two-stage MapReduce algorithm. The graph is represented by two input files, corresponding to the vertices (vector) and edges (matrix). In the first stage, the map phase transforms the input edges to set the destination vertex as the key. The following reduce phase applies a user-defined \emph{combine2} function on each group to produce partial values for each vertex. \emph{combine2} corresponds to a multiplication of a matrix element with a vector element. In the second MapReduce stage, the mapper is an identity mapper and the reducer encapsulates two user-defined functions, \emph{combineAll} and \emph{assign}. \emph{combineAll} corresponds to summing the partial multiplication results and \emph{assign} writes the new result in the vector. GIM-V can be used to express many iterative graph algorithms, such as PageRank, diameter estimation, and connected components.

HaLoop~\cite{HaLoop} supports efficient iterative computations on top of Hadoop~\cite{Hadoop}. It extends Hadoop with a caching and indexing mechanism, to avoid reloading iteration-invariant data and reduce communication costs. It also extends Hadoop’s API, offering a way to define loops and termination conditions.

Twister~\cite{Twister} extends the MapReduce API to support the development of iterative computations, including graph algorithms. It offers primitives for broadcast and scatter data transfers and implements a publish-subscribe protocol for message passing. 
 
\subsection{Dataflow}
Dataflow is a generalization of the MapReduce programming model, where a distributed application is represented by a Distributed Acyclic Graph (DAG) of operations. In the DAG, vertices correspond to data-parallel tasks and edges correspond to data \emph{flowing} from one task to another. As opposed to MapReduce, dataflow execution plans are more flexible and operators can support more than one input and output. In the DAG model, iterations can be supported by loop unrolling~\cite{Spark}, or by introducing complex iterate operators, as part of the execution DAG~\cite{spinning}. 

Spark~\cite{Spark}, Stratosphere~\cite{Stratosphere}, Apache Flink~\cite{Flink}, Hyracks~\cite{Hyracks}, Asterix~\cite{Asterix}, and Dryad~\cite{Dryad} are some of the general-purpose distributed execution engines implementing the DAG model for data-parallel analysis. Naiad~\cite{naiad} enriches the dataflow model with timestamps, representing logical points in computation. Using this timestamps, they build the \emph{timely dataflow} computational model, which supports efficient incremental computation and nested loops.

Dataflow systems offer different levels of abstraction for writing distributed applications. In Dryad and early versions of Stratosphere, the user describes the DAG by explicitly creating task vertices and communication edges.	User-defined functions are encapsulated in the vertices of the graph.
Modern dataflow systems, like Apache Spark and Apache Flink, offer declarative APIs for expressing distributed data analysis applications. Data sets are represented by an abstraction that handles partitioning across machines, like RDDs~\cite{RDDs}, and operators define data transformations, like map, group, join, sort.

It has been shown that the vertex-centric, scatter-gather, and other iterative models can be mapped to relational operations~\cite{spinning,GraphX,Pregelix,asymmetry}. For example, by representing the graph as two data sets corresponding to the vertices and edges, the vertex-centric model can be emulated by a join followed by a group-by operation. Algorithm~\ref{algo:pagerank-spark} shows an implementation of the PageRank algorithm in the Spark Scala API. The input edges ($links$) are grouped by the source ID to create an adjacency list per vertex and $ranks$ are initialized to 1.0.
Then, the $links$ are iteratively joined with the current $ranks$ to retrieve each node's neighbors' rank values. A reduce operation is applied on the neighbor ranks to compute the updated PageRank for each node. Spark, Flink, and AsterixDB, all currently offer high-level APIs and libraries for graph processing on their dataflow engines. Differential dataflow~\cite{differential-dataflow} also exposes a set of programming primitives, on top of which, higher-level programming models, such as vertex-centric, can be implemented.

\begin{algorithm}
\caption{PageRank in Apache Spark Scala API} \label{algo:pagerank-spark}
Input $lines$: a list of space-separated ID node pairs
\begin{algorithmic}
\State val links = lines.map$\{$ s =$>$
\State \hspace{\algorithmicindent} val parts = s.split(" ")(parts(0), parts(1))
\State $\}$.distinct().groupByKey().cache()
\State var ranks = links.mapValues(v =$>$ 1.0)
\State
\State for (i $<-$ 1 to iters) $\{$
\State \hspace{\algorithmicindent} val contribs = links.join(ranks).values.flatMap$\{$ case
\State \hspace{\algorithmicindent}\hspace{\algorithmicindent} (urls, rank) =$>$
\State \hspace{\algorithmicindent} \hspace{\algorithmicindent}\hspace{\algorithmicindent} val size = urls.size
\State \hspace{\algorithmicindent} \hspace{\algorithmicindent}\hspace{\algorithmicindent} urls.map(url =$>$ (url, rank / size))
\State \hspace{\algorithmicindent} $\}$
\State \hspace{\algorithmicindent} ranks = contribs.reduceByKey(\_ + \_)
\State \hspace{\algorithmicindent}\hspace{\algorithmicindent} .mapValues(0.15 + 0.85 * \_)
\State $\}$
\State val output = ranks.collect()
\end{algorithmic}
\end{algorithm}

\subsection{Linear Algebra Primitives}
\begin{figure}[!t]
\centering
\includegraphics[width=3.4in]{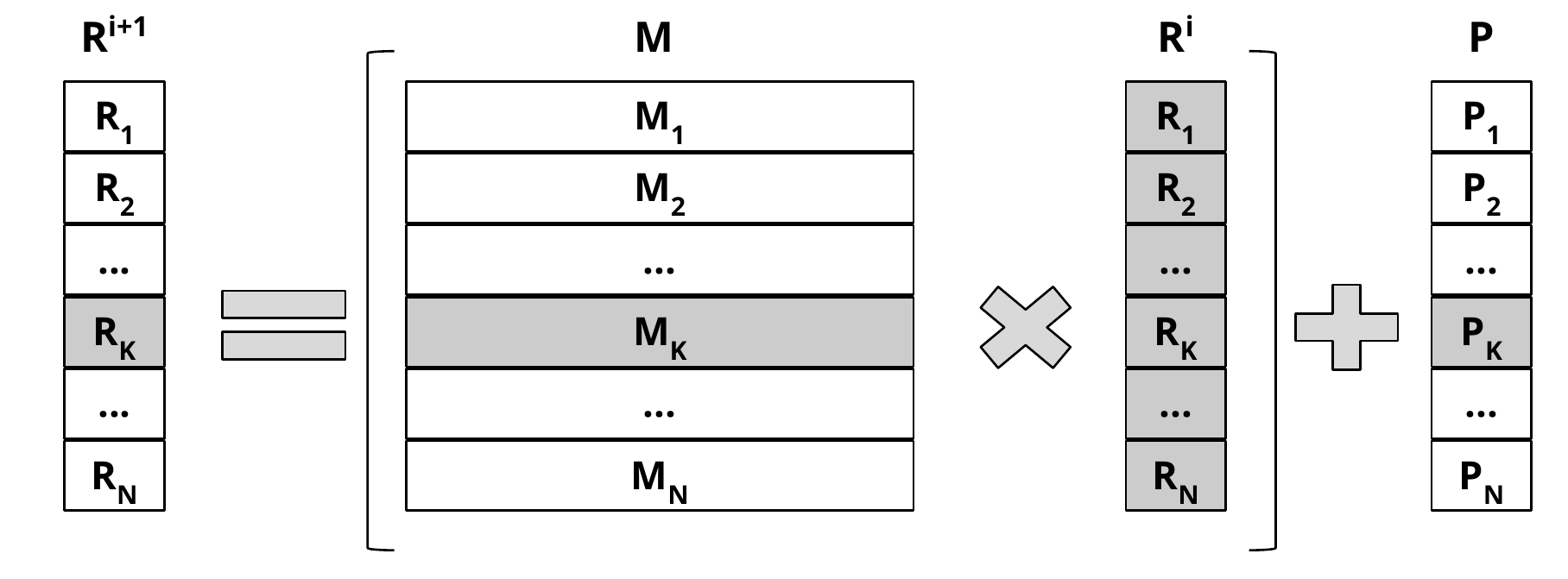}
\caption{PageRank computation as a vector-matrix multiplication. Vector $R_{i}$ represents the computed ranks at iteration $i$, $M$ is the input graph's adjacency matrix, with values scaled to account for the damping factor, and $P$ contains the transition probabilities.}
\label{fig:matrix-pagerank}
\end{figure}

Linear algebra primitives and operations have been long used to express graph algorithms. This model leverages the duality between a graph and its adjacency matrix representation~\cite{Linear-Algebra-Book}. A graph $G=(V, E)$ with $N$ vertices can be represented by a $NxN$ matrix $M$, where $M_{ij}=1$ if there is an edge $e_{ij}$ from node $i$ to node $j$ and 0 otherwise. Using this representation, many graph analysis algorithms can be expressed as a series of linear algebra operations. For example, a breadth-first search (BFS) starting from node $i$ can be easily expressed as matrix multiplication. Starting from an initial $1xN$ vector $y_{0}$, where only the $i_{th}$ element is non-zero, the multiplication $y_{1} = y_{0}*A$ will give the immediate neighbors of node $i$, $y_{2} = y_{1}*A$ will give the 2-hop neighbors of $i$, and so on. Figure~\ref{fig:matrix-pagerank} shows the computation of the PageRank algorithm in this model. Essentially, PageRank corresponds to the dominant eigenvector of the input graph's adjacency matrix. Starting from an initial vector of ranks $R_{0}$, PageRank can be computed by iteratively multiplying $R_{i}$ with the adjacency matrix and adding the transition probabilities to get $R_{i+1}$, until convergence.

Combinatorial BLAS~\cite{CombBLAS} and Presto~\cite{presto} are two seminal works in porting the linear algebra model for graph processing to a distributed setting. In Combinatorial BLAS, the graph adjacency matrix is represented as a distributed sparse matrix and graph operations are mapped to linear algebra operations between sparse matrices and vectors. The main operations perform matrix-matrix multiplication and matrix-vector multiplication and require user-defined functions for addition and multiplication. Presto extends the R language to support a distributed array abstraction, which can represent both dense and sparse matrices. Computation is distributed automatically based on the data partitioning. For instance, in Figure~\ref{fig:matrix-pagerank}, the shaded areas correspond to the data that is required for the computation of $R_{K}$. We notice that only a single row of the adjacency matrix is accessed and only one element of the transition probability vector. Thus, the computation task of $R_{K}$ should be sent to the partition that contains $M_{K}$ and  $P_{K}$. Extracting such access patterns is essential for partitioning data and computation in the linear algebra model.

\subsection{Datalog Extensions}
Datalog is a declarative logic programming language used as query language for deductive databases. Datalog programs consist of a set of \emph{rules}, which can be recursive. Datalog's support for recursions makes it suitable for expressing iterative graph algorithms. In~\cite{Socialite}, Datalog is enhanced with a set of extensions that allow users to define data distribution, efficient data structures for representing adjacency lists, and recursive aggregate functions, which can be efficiently evaluated using semi-naive evaluation~\cite{naive-evaluation}. Algorithm~\ref{algo:datalog-pagerank} shows the set of rules to compute an iteration of PageRank, using these extensions.

\begin{algorithm}
\caption{A PageRank iteration in SociaLite~\cite{Socialite}} \label{algo:datalog-pagerank}
\textbf{PageRank} (iteration i+1)
Input $N$: number of vertices
\begin{algorithmic}
\State EDGE (int $src$: 0..$N$, (int $sink$)).
\State EDGECOUNT (int $src$: 0..$N$, int $cnt$).
\State NODES (int $n$: 0..$N$).
\State RANK (int $iter$ : 0..10, (int $node$: 0..$N$, int $rank$)).
\State RANK($i$+1, $n$, \$SUM($r$)) :- NODES($n$), $r$ = 0.15/$N$;
\State\hspace{\algorithmicindent}\hspace{\algorithmicindent}\hspace{\algorithmicindent}\hspace{\algorithmicindent}\hspace{\algorithmicindent}\hspace{\algorithmicindent}\hspace{\algorithmicindent} \ \ :- RANK($i$, $p$, $r_{1}$), EDGE($p$, $n$),
\State\hspace{\algorithmicindent}\hspace{\algorithmicindent}\hspace{\algorithmicindent}\hspace{\algorithmicindent}\hspace{\algorithmicindent}\hspace{\algorithmicindent}\hspace{\algorithmicindent}\hspace{\algorithmicindent} EDGECOUNT($p$, $cnt$),
\State\hspace{\algorithmicindent}\hspace{\algorithmicindent}\hspace{\algorithmicindent}\hspace{\algorithmicindent}\hspace{\algorithmicindent}\hspace{\algorithmicindent}\hspace{\algorithmicindent}\hspace{\algorithmicindent} $cnt > 0$, $r = 0.85 * r_{1}/cnt$.
\end{algorithmic}
\end{algorithm}

A Datalog program is called \emph{stratifiable} if it contains no negation operations within a recursive cycle. For such a program, there exists a unique greatest fixed point for its rules. In order to parallelize Datalog programs, it is assumed that all input programs are stratifiable. To parallelize the recursive aggregate functions, these have to be \emph{monotone}, i.e. idempotent, commutative, and associative. Under these circumstance, it is shown in~\cite{Socialite} that delta stepping~\cite{delta-stepping} can be used to parallelize monotone recursive aggregate functions.

\subsection{Shared Partitioned Tables}
Distributed graph processing through shared partitioned tables is an idea implemented by Piccolo~\cite{piccolo}. The computation is expressed as a series of user-provided \emph{kernel} functions, which are executed in parallel and \emph{control} functions, which are executed on a single machine. Kernel instances communicate through shared distributed, mutable state. This state is represented as in-memory tables whose elements are stored in the memory of different compute nodes. Kernels can use a key-value table interface to read and write entries to these tables. The tables are partitioned across machines, according to a user-provided partitioning function. Users are responsible for handling synchronization and resolution functions for concurrent writes in the shared tables.

\section{Categorization of Distributed Graph Processing Systems}
\label{sec:categorization}
In this section, we present a taxonomy of recent distributed graph processing systems. Table~\ref{tbl:systems} contains 34 such systems and compares them in terms of supported graph programming abstractions, execution model, and communication mechanisms. We also review popular graph analysis applications, as these appear in recent distributed graph processing systems publications. We present the results in Table~\ref{tbl:applications}. 

\begin{table*}[]
\centering
\caption{Distributed graph processing systems comparison in terms of supported programming models, execution model, and communication mechanisms. $S$ stands for \emph{synchronous}, $A$ for \emph{asynchronous}, $H$ for \emph{hybrid}, and $I$ for \emph{incremental}.}
\label{tbl:systems}
\begin{tabular}{|p{3.2cm}||p{0.8cm}||p{3.3cm}|p{2cm}|p{2.8cm}|}
\hline
\textbf{System}    & \textbf{Year} & \textbf{Programming Model}  & \textbf{Execution} & \textbf{Communication} \\ \hline \hline
Pegasus~\cite{pegasus}            & 2009          & MapReduce                           & S  & Dataflow \\ \hline
Pregel~\cite{Pregel}             & 2010          & Vertex-Centric                      & S & Message-Passing \\ \hline
Signal/Collect~\cite{Signal-Collect}     & 2010          & Scatter-Gather                      & A, S & Message-Passing \\ \hline
HaLoop~\cite{HaLoop}             & 2010          & MapReduce                           & S    & Dataflow \\ \hline
Twister~\cite{Twister}            & 2010          & MapReduce                           & S     & Dataflow \\ \hline
Piccolo~\cite{piccolo}            & 2010          & Partitioned Tables                                   & S    & Shared global state \\ \hline
Apache Giraph~\cite{Giraph-VLDB}      & 2011          & Vertex-Centric                      & S  & Message-Passing \\ \hline
Comb. BLAS~\cite{CombBLAS} & 2011          & Linear Algebra     & S       & Message-Passing \\ \hline
Apache Hama~\cite{Hama}        & 2012          & Vertex-Centric                      & S     & Message-Passing \\ \hline
GraphLab~\cite{GraphLab}           & 2012          & Vertex-Centric                      & A, S     & Shared Memory \\ \hline
PowerGraph~\cite{PowerGraph}         & 2012          & GAS                                 & A, S       & Shared Memory \\ \hline
Giraph++~\cite{Giraph-Plus}           & 2013          & Subgraph-Centric                    & H   & Message-Passing  \\ \hline
Naiad~\cite{naiad}              & 2013          & Dataflow                            & A, S, I     & Dataflow \\ \hline
GPS~\cite{GPS}                & 2013          & Vertex-Centric                      & S    & Message-Passing \\ \hline
Mizan~\cite{Mizan}              & 2013          & Vertex-Centric                                   & S  & Message-Passing \\ \hline
Presto~\cite{presto}             & 2013          & Linear Algebra                                   & S   & Dataflow \\ \hline
Giraphx~\cite{giraphx}            & 2013          & Vertex-Centric                                   & A   & Shared Memory \\ \hline
X-Pregel~\cite{PregelX10}           & 2013          & Vertex-Centric                                   & S   & Message-Passing \\ \hline
LFGraph~\cite{LFGraph}            & 2013          & Vertex-Centric                                   & S  & Shared Memory \\ \hline
SociaLite~\cite{Socialite}          & 2013          & Datalog Extensions                                   & S   & Message-Passing \\ \hline
Trinity~\cite{Trinity}            & 2013          & Vertex-Centric                                   & A, S    & Message-Passing, Shared Memory \\ \hline
Graphx~\cite{GraphX}             & 2014          & GAS                                  & S     & Dataflow \\ \hline
GoFFish~\cite{GoFFish}             & 2014          & Subgraph-Centric                    & H  &  Message-Passing \\ \hline
Blogel~\cite{Blogel}             & 2014          & Vertex-Centric, Subgraph-Centric                    & H  &  Message-Passing\\ \hline
Seraph~\cite{seraph}             & 2014          & Vertex-Centric                                   & S  & Message-Passing \\ \hline
Cyclops~\cite{cyclops}            & 2014          & Vertex-Centric                      & S   & Message-Passing \\ \hline
GiraphUC~\cite{GiraphUC}           & 2015          & Vertex-Centric                      & A, H   & Message-Passing \\ \hline
Gelly~\cite{Gelly-post}              & 2015          & Vertex-Centric, Scatter-Gather, GAS & S   & Dataflow \\ \hline
Pregelix~\cite{Pregelix}           & 2015          & Vertex-Centric                      & S    & Dataflow \\ \hline
Apache Tinkerpop \cite{Tinkerpop}    & 2015          & Graph Traversals                    & S   & Message-Passing \\ \hline
PowerLyra~\cite{PowerLyra}          & 2015          & GAS                                 & S    & Shared Memory \\ \hline
NScale~\cite{NScale}             & 2015          & Neighborhood-Centric                & S   & Dataflow \\ \hline
Arabesque~\cite{Arabesque}          & 2015          & Filter-Process                      & S   & Message-Passing \\ \hline
Pregel+~\cite{pregel-plus}            & 2015             & Vertex-Centric    & S   & Message-Passing \\ \hline
\end{tabular}
\end{table*}

\subsection{Programming Model}
The vertex-centric model appears to be the most commonly implemented abstraction among the systems that we consider. Pregel, Apache Giraph, Apache Hama, GPS, Mizan, Giraphx, Seraph, GiraphUC, Pregel+, and Pregelix (Apache AsterixDB) implement the full semantics of the model. GraphLab, LFGraph, Cyclops, Gelly, and Trinity do not support graph mutations, while the former three do not support communication with vertices outside of the defined scope / neighborhood either. GraphX provides an operator called \emph{pregel} for iterative graph processing, but, in fact, this operator implements the GAS paradigm. Apache Tinkerpop provides connectors to Giraph and Spark, allowing the execution of graph traversals on top of their computation engines.

\subsection{Execution Model}
We encounter four execution techniques in the graph systems considered in this survey: synchronous (S), asynchronous (A), hybrid (H), and incremental (I).
Synchronous execution refers to implementations where a global barrier separates one iteration from the next. In such a model, during iteration $i$, vertices perform updates based on values computed in iteration $i-1$. On the other hand, in the asynchronous execution model, computation is performed on the most recent state of the graph. Synchronization can happen either through shared memory or through local barriers and distributed coordination. In a hybrid execution model, synchronous and asynchronous modes can coexist. For example, in Giraph++, computation and communication inside each partition happens asynchronously, while cross-partition computation requires global synchronization points. Incremental execution refers to the ability of a system to efficiently update the computation when its input changes, without halting and re-computing everything from scratch.

Synchronous execution is naturally the most common design choice. It simplifies application development and facilitates debugging. Asynchronous execution is usually supported together with synchronous or hybrid. The incremental execution model is only supported in Naiad.

\begin{table*}[]
\centering
\caption{Applications used in distributed graph processing systems papers to demonstrate programming models and evaluate performance.}
\label{tbl:applications}
\begin{tabular}{|p{5cm}||p{1.5cm}|p{6cm}|}
\hline
\multicolumn{1}{|c||}{\textbf{Application}} & \multicolumn{1}{c|}{\textbf{\#Appearances}} & \multicolumn{1}{c|}{\textbf{Programming Models}} \\ \hline
PageRank (and variations)                  & 31   & Vertex-Centric, Scatter-Gather, GAS, Dataflow, Linear Algebra, Subgraph-Centric, Partitioned Tables, MapReduce, Datalog Extensions \\ \hline
Shortest Paths (and variations)            & 15    & Vertex-Centric, Scatter-Gather, Dataflow, Linear Algebra, Subgraph-Centric, Datalog Extensions, GAS \\ \hline
Weakly Connected Components                & 12  & Vertex-Centric, GAS, Dataflow, Subgraph-Centric, MapReduce \\ \hline
Graph Coloring                             & 5  & Vertex-Centric, Scatter-Gather, Linear Algebra, GAS \\ \hline
Alternate Least Squares (ALS)        & 4  & Vertex-Centric, GAS, Linear Algebra \\ \hline
Triangle count                             & 4  & Vertex-Centric, Linear Algebra, Subgraph-Centric, Datalog Extensions  \\ \hline
K-Means                                    & 4  & Vertex-Centric, Linear Algebra, Subgraph-Centric, Partitioned Tables  \\ \hline
Minimum Spanning Forest / Tree             & 4 & Vertex-Centric \\ \hline
Belief Propagation (and variations)        & 2 & Vertex-Centric  \\ \hline
Strongly Connected Components              & 2  & Vertex-Centric, Dataflow \\ \hline
Label Propagation                          & 2 & Vertex-Centric  \\ \hline
Diameter Estimation                        & 2  & GAS, MapReduce  \\ \hline
Clustering Coefficient                     & 2  & Subgraph-Centric, Datalog Extensions  \\ \hline
Motif Counting                             & 2  &  Subgraph-Centric, Filter-Process  \\ \hline
BFS                                        & 2 &  Vertex-Centric, Subgraph-Centric \\ \hline
Centrality Measures                          & 1    & Linear Algebra \\ \hline
Markov Clustering                          & 1  & Linear Algebra \\ \hline
Find Mutual Neighbors                      & 1 & Datalog Extensions \\ \hline
SALSA                                      & 1  & Dataflow \\ \hline
n-body                                     & 1  & Partitioned Tables \\ \hline
Bipartite Matching                         & 1  & Vertex-Centric \\ \hline
Semi-Clustering                            & 1 & Vertex-Centric \\ \hline
Random Walk                                & 1   & Vertex-Centric \\ \hline
K-Core                                     & 1 & Vertex-Centric \\ \hline
Approximate Max. Weight Matching        & 1 & Vertex-Centric \\ \hline
Graph Coarsening                           & 1  &  Subgraph-Centric \\ \hline
Identifying Weak Ties                      & 1   & Subgraph-Centric \\ \hline
Frequent Subgraph Mining                   & 1 & Filter-Process \\ \hline
Finding Cliques                            & 1   & Filter-Process \\ \hline
\end{tabular}
\end{table*}

\subsection{Communication Mechanisms}
We come across four different communication mechanisms. In the Message-Passing model, the state is partitioned across worker tasks and updates to non-local state happen by sending and receiving messages. Worker tasks have read-write access to local state but they cannot directly access and modify state on a different machine. On the contrary, the shared memory mechanism allows tasks in different machines to communicate by mutating shared state. Systems that employ this mechanism need to account for race conditions and data consistency. In the dataflow model, operators are usually stateless and data flows from one stage of computation to the next. In order to efficiently support graph computations in this model, dataflow systems offer explicit or automatic caching mechanisms. For example, Spark has a \emph{cache()} method, which can be used to cache the graph structure, which is static. In Stratosphere and Flink, the optimizer will detect loop-invariant data and cache them automatically.

\subsection{Applications}
To further assess distributed graph programming model expressiveness and systems usability, we survey the applications which appear in recent distributed graph processing systems papers. We gather and group graph algorithms that are used in the papers introducing the systems of Table~\ref{tbl:systems}. We choose applications that appear as examples or pseudocode to demonstrate APIs and programming model interfaces, as well as applications used for performance evaluation. We base this study on the assumption that paper and systems authors choose \emph{representative} algorithms to include in their papers, in order to show their systems' applicability and efficiency. Table~\ref{tbl:applications} shows the most commonly encountered applications, sorted by appearance frequency. For each application, we also list the programming models in which they are implemented.

Unsurprisingly, we find the PageRank algorithm to be extremely popular. This algorithm appears in 31 out of the 34 examined systems and we encounter implementations in 9 out of the 11 programming abstractions that we present in Section~\ref{sec:abstractions}. Shortest paths calculation and weakly connected components appear in more than one third of the papers. Graph coloring, ALS, k-means clustering, and minimum spanning forest also appear to be quite commonly used. However, we notice that the majority of applications are only encountered once or twice. This is partially explained by the fact that some of them, like finding cliques, serve the purpose of introducing a specialized programming model, like filter-process. Nevertheless, it appears that the vertex-centric model has been used to implement most of the applications in the table.

\section{Discussion and Open Issues}
\label{sec:open}
Similarly to how the introduction of the map-reduce programming model simplified large-scale data analysis to a large extend, the invention of the vertex-centric model revolutionized the area of distributed graph processing. \emph{Thinking like a vertex} has proved to be a valuable abstraction that allows writing comprehensive programs for a variety of graph problems. However, despite its wide adoption, researchers and users have also identified many of its shortcomings~\cite{pregel-salihoglu,pregel-plus,Giraph-Plus}. The understanding of the abstraction's limits has spawned novel specialized models, such as the neighborhood-centric model for ego-network analysis and the filter-process model for graph mining tasks. At the same time, lower-level abstractions, such as the partition-centric model, are proposed as alternatives for better performance.

Our study suggests that no single model is suitable for all classes of graph algorithms. It is an open challenge for researchers and systems designers to either invent a more expressive and flexible programming model or explore the possibility of supporting multiple programming abstractions on top of the same platform. To that end, general-purpose dataflow systems look promising. It has already been shown how the vertex-centric, scatter-gather, and GAS abstractions can be mapped to relational execution plans, and how distributed dataflow frameworks can efficiently support them~\cite{spinning,Pregelix,GraphX}. Nevertheless, current systems still rely on the user for efficient implementations. The research community could investigate improving existing relational optimizers for handling graph tasks or building graph processing optimizers that could choose the most suitable model based on the application characteristics and input graph properties.

A worrying trend in recent graph processing literature is the lack of diversity in applications used for evaluating usability and performance. A very small set of graph algorithms (PageRank, Connected Components, Shortest Paths) is being repeatedly used and promoted as a representative benchmark, when, in essence, these algorithms are quite similar; single-stage, value-propagation iterative algorithms. We believe that the research community would benefit immensely from exploring more complex, multi-stage algorithms to further challenge the expressiveness of existing abstractions and the performance of current systems implementations.

Another open question in the area of distributed graph processing clarifying the conditions that make distribution necessary. It has been recently shown that single-thread implementations can outperform some distributed graph processing systems~\cite{scalability-cost}. At the same time, significant progress has been made in building efficient single-machine graph processors that can handle billion-edge graphs~\cite{graphchi,x-stream,chaos}. Thus, it is critical to quantify parallelization and communication overheads in distributed graph processing systems and clarify the circumstances under which a distributed implementation would be preferable over a centralized solution.

Finally, a largely unexplored direction in the area of distributed graph processing is support for dynamic graphs and temporal analytics. Most graphs are highly dynamic in nature or are generated in real-time from streaming sources. However, most of the existing graph processors assume a static immutable graph structure. Systems and models need to evolve in order to facilitate real-time graph analytics and incremental processing of changing graph inputs. Kineograph~\cite{Kineograph} and Chronos~\cite{Chronos} are two pioneering works in this direction. Kineograph supports incremental graph processing through consistent periodical snapshots, while Chronos proposes a novel in-memory layout of temporal graphs that allows for efficient computation across a series of snapshots.

\section{Related Work}
\label{sec:related}
The work that is closest to ours is a recent survey of vertex-centric frameworks for graph processing~\cite{vertex-centric-survey}. It presents a extensive study of frameworks that implement the vertex-centric programming model and compares them in terms of system design characteristics, such as scheduling, partitioning, fault-tolerance, and scalability. Moreover, it briefly introduces subgraph-centric frameworks, as an optimization to vertex-centric implementations. While there is some overlap between this work and ours, the objective of our study is to present the first comprehensive comparison of distributed graph processing abstractions, regardless of the specifics of their implementations. While in~\cite{vertex-centric-survey} the discussion revolves around certain frameworks, in our work, we first consider the programming models decoupled from the systems, then build a taxonomy of systems based on their implemented model.

A notable study of parallel graph processing systems is presented in~\cite{delft-tech-report}. The work surveys over 80 systems, spanning single-machine, shared-memory, and distributed architectures. The scope of their study is much wider than ours, yet their results are driven by system implementations and platform design choices. With regard to programming models, they consider general-purpose processing systems, vertex-centric, and subgraph-centric, but do not further expand on execution semantics, user-facing interfaces or limitations.

While the focus of our study is to describe and compare available programming abstractions for distributed graph processing, several recent studies focus on the performance of modern distributed graph processing platforms. In~\cite{Study-Delft}, the authors compare the performance of six systems, including general-purpose data processing frameworks, specialized graph processors and a non-distributed graph database. The study evaluates these systems across a number of performance metrics, such as execution time, CPU and memory utilization, and scalability, using real-world datasets and a diverse set of graph algorithms. Similarly,~\cite{experimental-study-vldb} presents an experimental evaluation of distributed graph processing platforms, but covering only specialized graph processors. The also study the effectiveness of implemented optimization techniques and compare performance with a single machine system as baseline. A similar but more extensive study can be found in~\cite{experimental-survey-springer}. Both these studies conclude that no single system performs best in all cases and also highlight the need for a standard benchmark solution, that could simplify the performance comparison of graph processing platforms. Even though a standard benchmarking solution for graph processing systems does not exist yet, ~\cite{benchmark-vision,Graphalytics} are two significant steps towards that direction. They propose a clear vision for a benchmark, listing challenges that need to be addressed and share early results in designing a graph processing benchmark, called Graphalytics. Graphalytics supports graph generation with custom degree distributions and structural characteristics and already supports several systems and algorithms.

\section{Conclusion}
\label{sec:conclusion}
Graphs are practical data structures that can elegantly capture relationships between data items. Efficiently analyzing large-scale graphs is gradually becoming an essential requirement for business intelligence, social networks, web applications, and modern science. In order to cope with the rapid increase of graph data sizes, distributed graph processing has proven to be a popular solution. To facilitate the development of graph algorithms in a distributed environment, several high-level abstractions for graph processing have been proposed in recent literature.

In this survey, we present a comprehensive review of the most prevalent high-level abstractions for distributed graph processing. We analyze and compare their semantics and user-facing interfaces. We comment on their usability and expressiveness, identifying representative applications and computation patterns that are hard to express. We further survey proposed performance optimizations and model variations. We then look at recent distributed graph processing systems and categorize them in terms of programming model, execution model, and communication mechanisms. We also survey graph analysis applications that are used in recent papers to demonstrate systems usability and performance. Finally, we discuss open challenges in the area of distributed graph processing and identify future research directions. We believe that the community would benefit from experimenting with more complex applications and exploring the limits of current abstractions. We see a promising direction in supporting multiple graph programming models on top of general-purpose dataflow systems and we anticipate a great research potential in the area of real-time and temporal graph analytics.

\end{document}